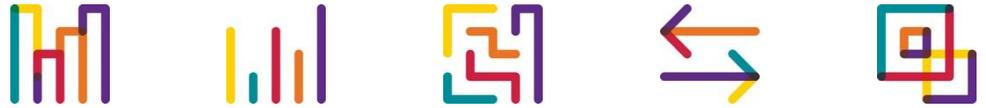

# NYC Recovery at a Glance – The Rise of Buses and Micromobility

Suzana Duran Bernardes, Zilin Bian, Siva Sooryaa Muruga Thambiran, Jingqin Gao, Chaekuk Na, Fan Zuo, Nick Hudanich, Abhinav Bhattacharyya Kaan Ozbay, Shri Iyer, Joseph Y.J. Chow, Hani Nassif

Contact: c2smart@nyu.edu
c2smart.engineering.nyu.edu

## Executive Summary

New York City (NYC) is entering Phase 4 of the state's reopening plan, starting July 20, 2020. This white paper updates travel trends observed during the first three reopening phases and highlights the spatial distributions in terms of bus speeds and Citi Bike trips, and further investigates the role of micromobility in the pandemic response.

## Key Findings

- Vehicular traffic volumes (1) were restored to 80% of pre-pandemic levels by the week of July 6, 2020. Despite the increased vehicular volumes, speeding remains a problem in the city, with a 57% increase in school zone speeding tickets observed over May-June (4), compared with pre-pandemic levels in 2020. Travel times on the 495 Corridor (2) remained low during April to June but slightly increased by an average of 16% eastbound and 11% westbound in the first week of July, compared with the same week in April. However, these travel times are still about 28% lower in both directions compared to pre-pandemic levels.
- Data from July 2020 (1) shows that NYC average subway and commuter rail ridership continues to lag, being 80% down in the first week of July compare to 2019. Estimated bus ridership and scheduled Access-A-Ride trips (1) have recovered faster, down only 50% in the first week of July compare to 2019. Meanwhile average daily bus ridership reached 866,297 in June, 9% higher than subway ridership, according to MTA data.
- Weigh-in-motion (WIM) data from C2SMART's testbed on the Brooklyn-Queens Expressway (BQE) (5) showed that average daily traffic (ADT) is down by only 6% for both Queensbound (QB), and Staten Island-bound (SIB) traffic in June, as compared with February 2020. Average daily truck traffic (ADTT) for SIB is down by only 1% in June, and is 4% higher QB in June compared to February. The average vehicle speed on BQE has decreased gradually but is still 12% higher for SIB and 1% higher for QB in June, as compared to February.
- MTA buses zonal speeds averaged 30% higher in the third week of April 2020 from 8 AM to 6 PM, compared with the same month of 2019, based on data aggregated from MTA Bus Time (7). Mean bus speeds dropped by 11% by the third week of June but are 16% higher than the corresponding speed from April 2019. However, bus speeds have not reduced uniformly across the city.
- The daily number of Citi Bike trips (8) began increasing by mid-April with peak ridership days, matching days with higher temperatures, which may indicate recreational trips. Higher trip density is observed near some hospitals and parks during March to June.
- Based on the deep-learning based video-processing algorithm (6) introduced in the previous issue of whitepaper, a slight increase in average pedestrian density (+8%) was observed from representative weekdays at 11 locations in NYC. The average percentage of pedestrians following social distancing guidelines of at least 6 feet of distance between others at select locations was estimated to be 82% in the last two weeks of June.
- 24-hour temporal density distribution shows that the cyclist density has approached pre-pandemic levels at 5th Avenue and 42nd Street on June 24. Meanwhile, an increase in car and pedestrian density has also been observed in the end of June compare to April-May.

## Mobility Trends

According to MTA data (1), average subway and commuter rail ridership in NYC continues to stay low – a 78%, 78% and 81% decline in average daily ridership was observed for NYC subway, Long Island Rail Road (LIRR) and Metro North Railroad (MNR) in the week of July 6 compare to daily average in 2019. However, the average daily vehicular traffic via MTA bridges and tunnels (1) has been restored to 80% of pre-pandemic levels by the week of July 6. Bus ridership (1), estimated from models that use MetroCard and OMNY swipes and taps and Automatic Passenger Counter (APC) data that is available on a portion of the MTA bus fleet, has been restored to 55% or pre-pandemic levels, and scheduled Access-A-Ride trips (1) restored to 53% in the week of July 6 compare to 2019 weekday/Saturday/Sunday average. Figure 1 shows the ridership trends reported by MTA compared to 2019 weekday/Saturday/Sunday average. According to the MTA data, the relative drop in bus and commuter rail ridership is slightly lower on weekends in 2020 than in 2019, while Access-A-Ride trips experience a relatively greater drop on weekends, possibly due changes in discretionary trip patterns.

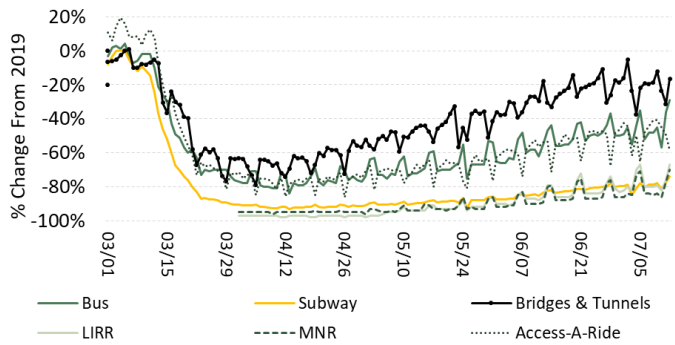

**Figure 1** Mobility trends in NYC compared to 2019 Weekday/Saturday/Sunday average, Source: MTA (1)

Travel times across the 495 corridor (2) connecting the NJ highway network to the Lincoln Tunnel, across Manhattan-34th Street to the Queens-Midtown Tunnel into Long Island, were up by an average of 16% eastbound and 11% westbound in the first week of July 2020, compared with the same week in April 2020. However, these travel times are still about 28% lower in both directions compared to pre-pandemic levels (first week of March). School zone speeding tickets (4) were up 57% from May 17 to June 14 compared to January 1 to March 12 of 2020. Although the number of reported motor vehicle collisions (4) was still 62% lower in June compare to same month in 2019, the percentage of cyclist injuries among all injuries has increased from 11% to 14% in the last two weeks of June.

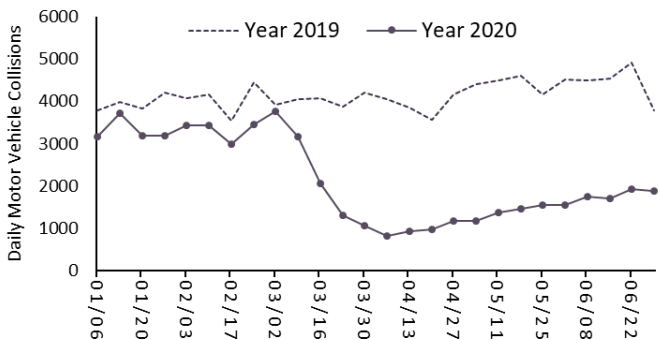

**Figure 2** Reported motor vehicle collisions, Source: (3)

As of June 28, Weigh-in-motion (WIM) data from C2SMART's testbed on the Brooklyn-Queens Expressway (BQE) (5) showed that average daily traffic is down by only 6% for both Queensbound (QB), and Staten Island-bound (SIB) in June, as compared with February 2020. Average daily truck traffic for SIB has almost rebounded to pre-pandemic levels, with 1% decrease in June, and has increased by 4% in the QB direction versus pre-pandemic levels. The average vehicle speed through this section of the BQE has been decreasing gradually as volumes return but is still 12% higher for SIB and 1% higher for QB in June 2020, as compare to February 2020.

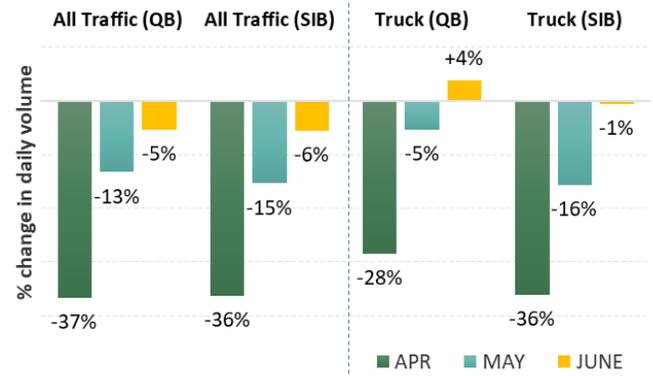

**Figure 3** Average daily traffic (ADT) and average daily truck traffic (ADTT) from BQE WIM Installation, Source: (5)

### Crowding and Density on City Streets

The previous issue of this white paper introduced an efficient, artificial intelligence (AI)-driven, continuous, real-time social-distancing big data acquisition framework that leverages traffic cameras along with a deep learning-based video processing method for analyzing travel patterns at the local level without the need for new equipment or tools (6). Object detection is being applied to real-time traffic camera videos at multiple key locations within NYC to provide a holistic view of evolving crowd density and social distancing patterns during the COVID-19 outbreak and the subsequent recovery process, while also observing timeseries travel volumes for different modes. A slight increase in average pedestrian density (+8%) has been observed from 11 locations in NYC on June 24 as compared to May 27, which are representative weekdays that both featured alike camera conditions. The average percentage of pedestrians following social distancing guidelines of remaining a minimum of 6 feet from others was observed at 82% during the last two weeks of June.

Figure 4 illustrates the 24-hour temporal density of daily snapshots of the distributions of pedestrian, cars, and cyclists at a selected location (5 Ave/42 St, Manhattan). A gradual increase in car and pedestrian densities are observed from April to June. In June 24, the cyclist density has approached pre-pandemic levels at this location.

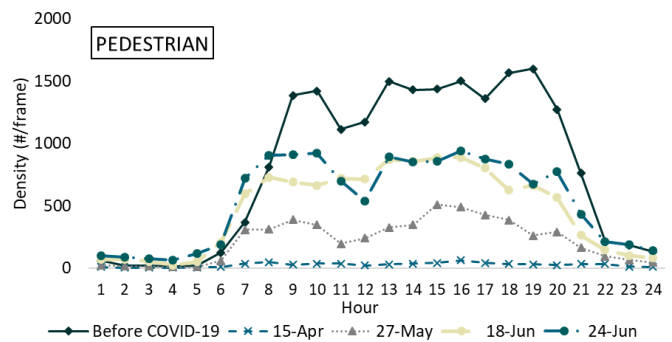

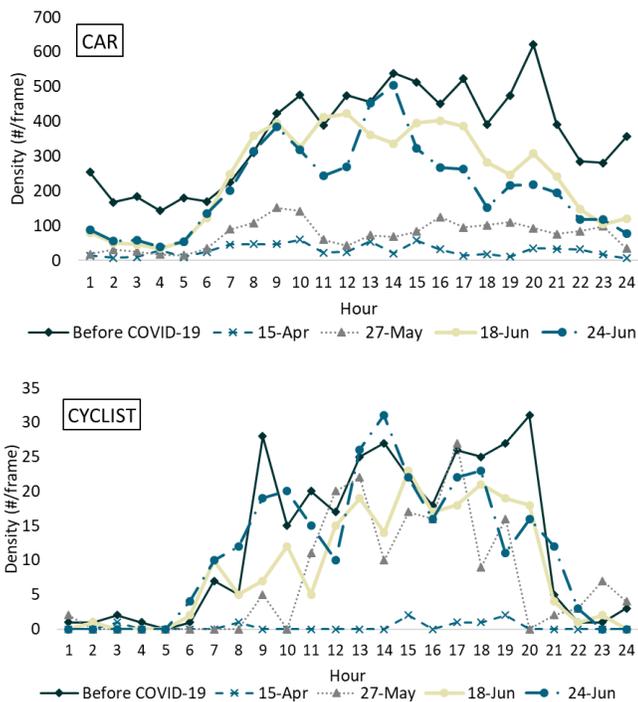

**Figure 4** Temporal Distributions of Pedestrian, Car, and Cyclist Densities (5 Ave/42 St, Manhattan)

## New York City Bus Ridership Is on the Rise

New York City bus system has increasingly become a crucial travel mode during the city's recovery. According to MTA data, Average daily ridership has overtaken subway ridership during the pandemic since April.

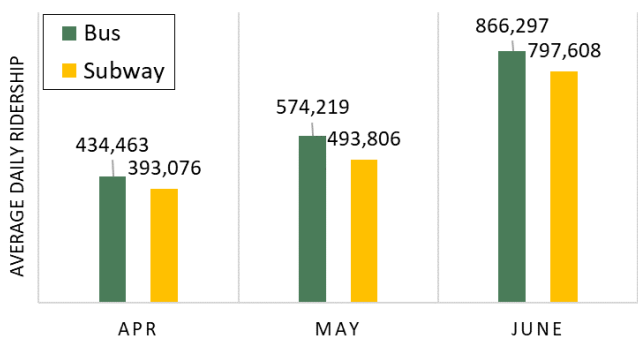

**Figure 5** MTA subway and bus average daily ridership, Source: MTA (1)

## Citywide Bus Speed

Largely due to overall lower traffic volumes on city streets and other policy and planning actions, buses have experienced a resurgence in speeds. Table 1 and Figure 6 compare bus traffic speeds extracted from MTA Bus Time API (5) from four time periods: the third weeks of April 2019, June 2019, April 2020 and June 2020. Bus speeds were calculated every 30 seconds for each bus based on its space mean speed between two consecutive GPS points using MTA Bus Time data and then aggregated to NTAs. Average zonal speed is then calculated for 8 AM to 6PM on all weekdays; weekends were excluded. Using April 2019 as baseline, although higher mean bus speeds are observed in April and June 2020 compared with 2019, they are also more dispersed (wider distributions).

**Table 1** Comparison of NYC average zonal bus speeds from the 3rd weeks of three months using April 2019 as baseline

| Citywide NTA Zonal Bus Speed 8 AM-6 PM (% change) | | | | |
|---|---|---|---|---|
| **Time Period** | **Mean** | **Min** | **Max** | **Std** |
| **June-2019** | -1% | 4% | -4% | -1% |
| **Apr-2020** | +30% | +72% | +19% | +28% |
| **June-2020** | +16% | +48% | +16% | +33% |

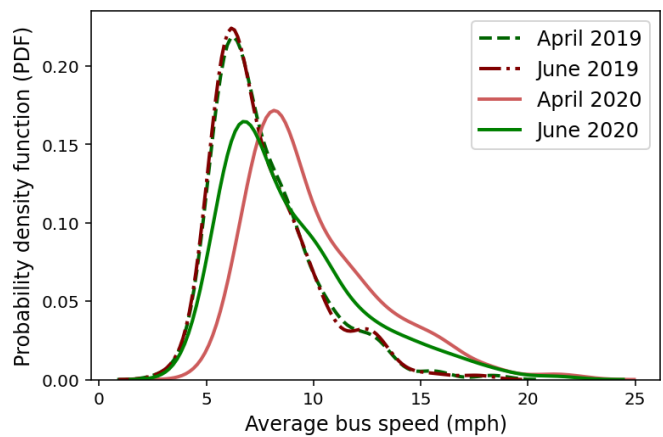

**Figure 6** Probability density function of average daily bus speed (8 AM to 6 PM)

April 2020, the height of observed traffic drop due to the pandemic, has a higher zonal average speed, maximum speed, and a lower standard deviation compared to June 2020. By June 2020, mean zonal bus speed dropped by 11% versus April 2020, but is still 16% higher than June 2019. The maximum zonal speed value is similar to April's number while the minimum value is lower; this might be due to the possibility that bus speeds haven't reduced uniformly across the city. It is worth noting that besides the impact of the pandemic, the opening up of new bus lanes on 14th Street, Lexington Avenue and areas in the Bronx could have caused the increase in bus speeds.

## Spatial Distribution of Bus Speeds

The maps displayed in Figure 7 (a-d) show the average daytime zonal bus speeds in each neighborhood tabulation area (NTA), recorded in April/June 2019 and 2020, based on Bus Time data provided by the MTA. NTAs were created to project populations at a small area level. The darker colored zones represent NTAs where on average, there are lower speeds recorded while the lighter colored zones have higher speeds on average.

The relative differences in bus speeds for each NTA in June 2020 is also compared with the same month in 2019 in Figure 8. Lower and Midtown Manhattan, Staten Island, and outer parts of the city but also NTAs such as Flushing and Forest Hills are the zones that the greatest increases in average bus speeds in June 2020 compared with 2019. Again, this may be due to other measures besides the impact of the pandemic such as new bus-only lanes.

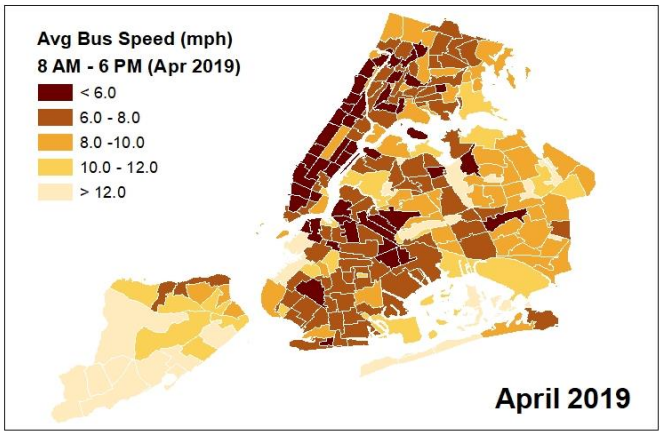

(a) Average bus speeds in April 2019

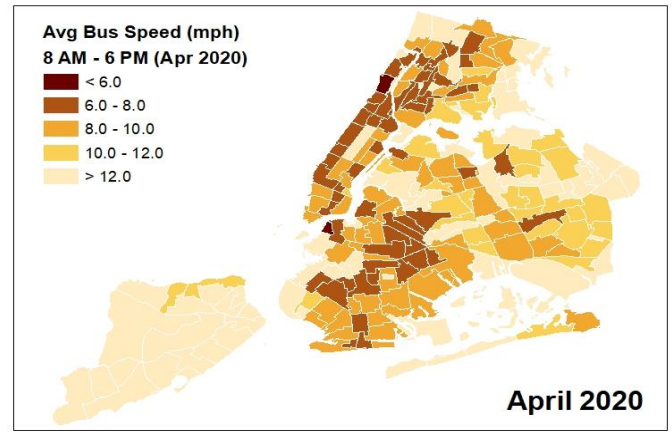

(b) Average bus speeds in April 2020

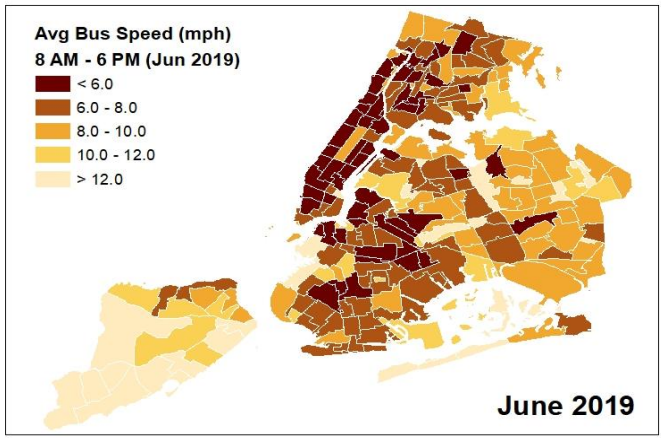

(a) Average bus speed in June 2019

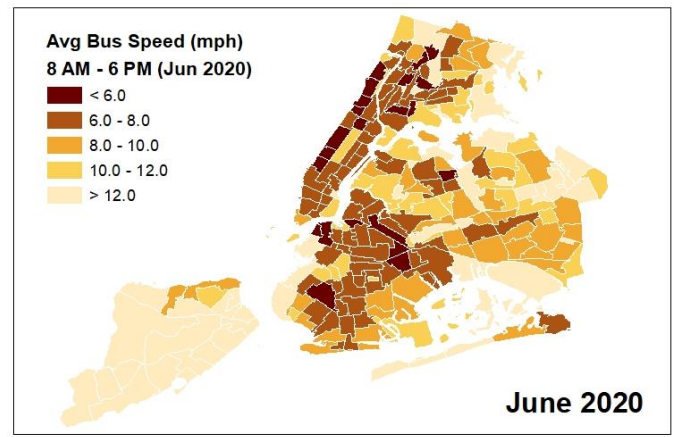

(c) Average bus speeds in June 2020

**Figure 7** Average bus speeds in New York City NTAs in April & June 2019-2020 (8AM to 6PM)

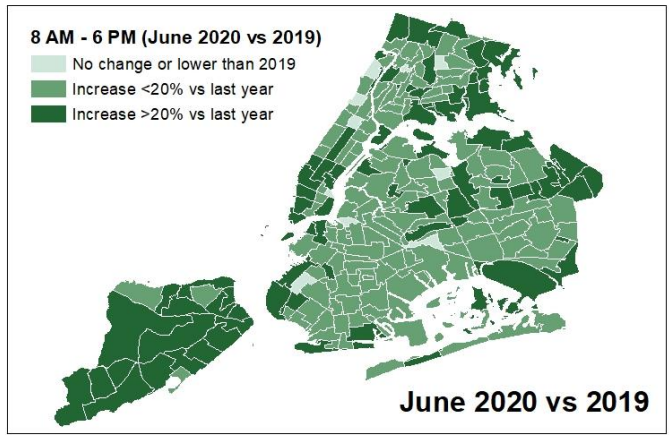

**Figure 8** Percentage difference in average bus speed in June 2019 and 2020 (8AM to 6PM)

Bus Speed During Pandemic and After Reopening

The NTAs that experienced the highest difference in speed from April to June 2020 are highlighted in Table 2. This may indicate a faster recovery of local activities in these zones as most are social and economic centers of influence that have many stops.

**Table 2** NTAs with the highest differential decrease in speed numbers from April 2020 to June 2020 (8AM to 6PM)

| NTA | Borough | (%) decrease |
|---|---|---|
| Williamsburg | Brooklyn | 28.1 |
| Prospect Heights | Brooklyn | 27.8 |
| Lenox Hill-Roosevelt Island | Manhattan | 26.8 |
| East Harlem South | Manhattan | 26.4 |
| SoHo-TriBeCa | Manhattan | 26.2 |

**Continued Rise of Micromobility**

Citi Bike

Based on Citi Bike system data (8), Citi Bike trips increased by more than 40% in the first week of March 2020, just before stay-at-home orders were issued, when compared to the same period last year. After the stay-at-order was issued, the number of daily trips experienced a significant drop. Figure 9 shows the percentage difference between the total number of rides per week for 2020 and 2019.

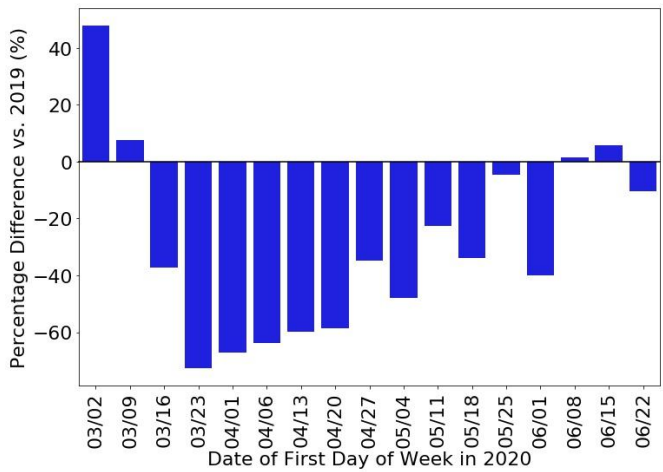

**Figure 9** Weekly percentage difference between the total number of Citi Bike rides in 2020 and 2019, Source: (8)

The daily number of trips starts to increase again by the middle to end of April with peaks in the beginning of May, as shown in Figure 10. These peaks in the number of trips in May match peaks in temperature for the same period, which might indicate that the increase in rides are correlated not only to the developments of the pandemic, but also to the weather. Another observation from Figure 10 is that the pattern of constant increase in rides started around the same time that MTA announced that subways would stop running overnight.

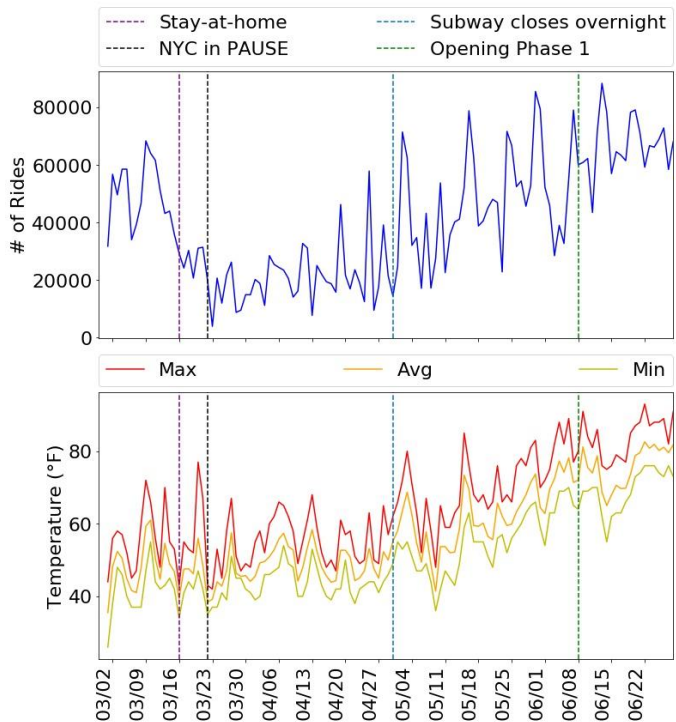

**Figure 10** Time series of the number of Citi Bike rides and temperature per day

Density maps presented in Figure 11 show the concentration of rides ending at each Citi Bike station, or destinations. The number of trips were weighted to fit a range of 0 to 1 using the 2019 range to show areas with the greatest change in concentration. When looking into how these trips are distributed geographically in New York City during the stay-at-home period (Figure 11 (a)), the total number of trips to different destinations is lower than 2019 but more well distributed throughout the city with some hotspots near hospitals and parks. In 2019, for the same period, the destinations are mainly concentrated in two areas of lower and midtown Manhattan (Figure 11 (b)).

Although midtown Manhattan has begun to see a similar concentration of rides as June 2019, the overall distribution of rides throughout the city still differs from the same period in 2019 (Figure 11 (c) and (d)). Hotspots close to hospitals and parks, especially Central Park and parks along the west side, have emerged. The increase in stations closer to parks in both years might be correlated to the increase in temperatures for the period, but they are more distributed into new areas such as the Upper East Side in 2020. New stations were also installed in 2020 in Upper Manhattan and in the Bronx, which might lead to different observation in behavior.

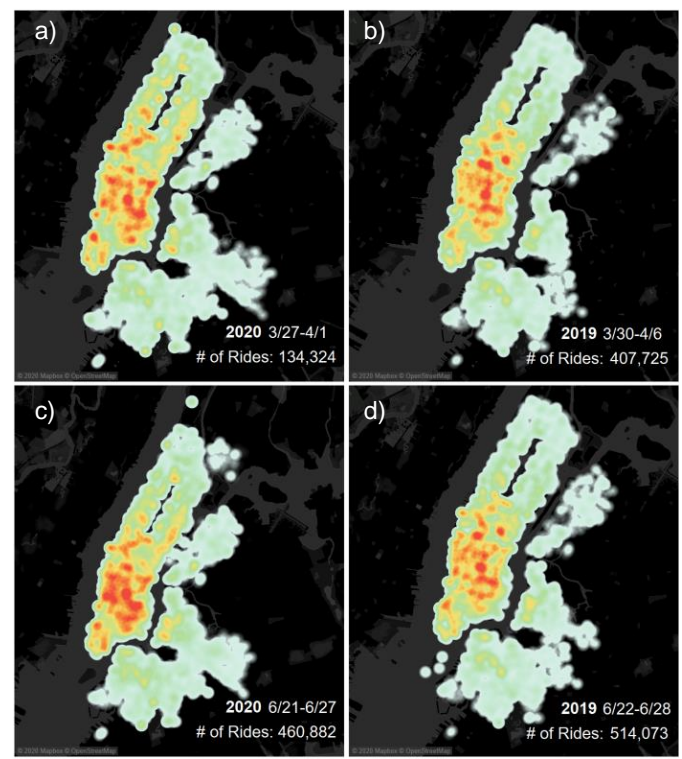

**Figure 11** Density map of total number of Citi Bike rides to destinations for the 14th week for 2020 (a) (3/27 - 4/1) and 2019 (b) (3/30 - 4/6) and for the 26th week for 2020 (c) (6/21 - 6/27) and 2019 (d) (6/22 - 6/28)

From April to June 2020, the 1 Ave & E 68 St Citi Bike station registered the highest (139 trips in April) and second highest (257 trips in May and 355 trips in June) average number of trips daily. This station is the second closest to the New York-Presbyterian hospital, one of the top-ranked hospitals in the U.S. and highly hit by COVID-19 (11). The only station that was in the top 5 destinations for all the pandemic months was located at West St & Chambers St. This station is located in the richest zip-code (13) in New York City with lots of green area, which might indicate that some of the rides to this destination were for recreational purposes.

The average duration of Citi Bike trips has also changed since the beginning of March, as shown in Figure 12. For the period before the pandemic, the distribution of trip durations followed a similar pattern for the same period of last year. However, the distribution during the pandemic shows that Citibikers are taking longer rides, indicating that people are using bikes as an alternative mode for trips that would have used another mode prior to the pandemic. In addition, a survey performed by Trek corporation (13) indicates that 85% of the respondents perceived cycling as safer than transit for social distancing.

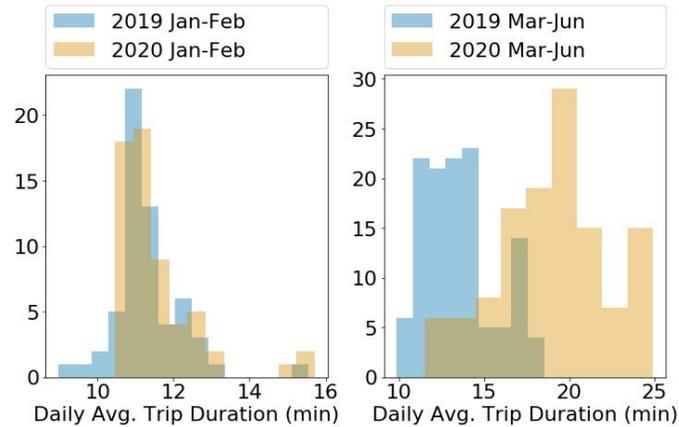

**Figure 12** Distribution of daily average Citi Bike trip duration in minutes

The daily average number of rides for each month in which the duration was higher than 45 minutes, the maximum allowable for each single ride for annual membership holders without an additional fee, is shown in Table 3. From Table 3, it is possible to notice that more people might be willing to pay the extra cost in the months in which the city started to open and the temperatures started to increase.

**Table 3** Daily average number of rides between 45 and 120 minutes

| Month | Daily Avg. Number of Rides >45 mins |
|---|---|
| March | 977 |
| April | 1,340 |
| May | 3,520 |
| June | 4,148 |

### Electric bike, scooter, and moped

In the first week of April 2020, New York State began permitting electric bicycles, scooters, and mopeds. Even though some micromobility services were halted or reduced during the pandemic (about 74% North American Shared Mobility service are in operation as of July 2020 (9)), with social distancing practices in effect, they provide an alternative mode of travel.

While NYC ridership data has not been released yet, fleet operators like Gotcha, which offers electric bike rentals, electric scooter rentals and ride sharing in about 35 cities across the U.S., reported spikes in the number of rides in many of its markets. For example, a 185% increase in daily trips was observed in Mobile, AL since April (14). Scooter company Spin also saw a 34% average increase in new daily active users since April and a 44% increase in trip duration in May (15).

Moped-rental service Revel saw an initial drop in its ridership with the start of stay-at-home orders in NYC. According to Revel data from its communications team (16), average daily rides fell from over 4,100 per day in the first half of March to under 1,850 per day in the last half of March. However, ridership rebounded beginning in April and reached 8,800 average rides per day by May and close to 18,000 by June in NYC.

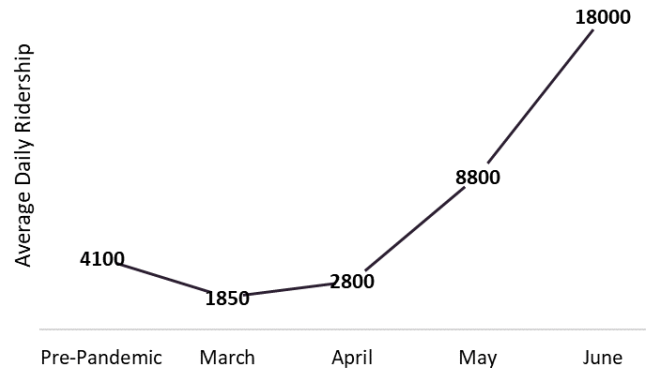

**Figure 13** Average Daily Revel Ridership in NYC (March-June 2020, March-May data is based on the second half of the month), Source: Revel (16)

### Summary of Findings

As NYC is on track for the fourth phase of reopening following the pandemic, traffic volumes continue to approach pre-pandemic levels. Other modes, such as bus and micromobility, are on the rise and have even surpasses pre-pandemic volumes in some cases. These modes are being increasingly counted on as an alternative to the subway, as safer and less-crowded travel options. The quick adoption of electric bikes, scooters, and mopeds by several cities in the last few years show that these modes are economical and sustainable options of moving people and goods.

This paper reflects the Center's perspective as of July 22, 2020 based on data collected in June 2020. C2SMART researchers are continuing to collect data and monitor both mobility and sociability trends and regularly update findings during the reopening of the New York City.

*Note: all data is preliminary and subject to change. The authors gratefully acknowledge Revel for sharing its ridership data in NYC. For more information please contact c2smart@nyu.edu. For more data and visualizations, please visit our COVID-19 Dashboard:* c2smart.engineering.nyu.edu/covid-19-dashboard/